# EQUILIBRIUM STATISTICAL MECHANICS OF FRUSTRATED SPIN GLASSES: A SURVEY OF MATHEMATICAL RESULTS[1]

Dimitri PETRITIS

Institut de Recherche Mathématique

Université de Rennes I and CNRS URA 305

Campus de Beaulieu

F - 35042 Rennes Cedex

petritis@levy.univ-rennes1.fr

23 November 1994

## Abstract

After a rapid introduction to the physical motivations and a succinct presentation of heuristic results, this survey summarises the main mathematical results known on the Edwards-Anderson and the Sherrington-Kirkpatrick models of spin glasses. Although not complete proofs but rather sketches of the relevant steps and important ideas are given, only results for which complete proofs are known — and for which the author has been able to reproduce all the intermediate logical steps — are presented in the sections entitled 'mathematical results'. This paper is intended to both physicists, interested to know which articles among the multitude of papers published on the subject go beyond the heuristic arguments to obtain rigorous irrefutable results, but also to the mathematicians, interested in finding out how rich is the physical intuitive way of thinking and in being inspired by the heuristic results in view of a mathematical rigorisation. An extended, but not exhaustive, bibliography is included.

---

[1]Work partially supported by EU grant CHRX-CT93-0411.

1991 *Mathematics Subject Classification*: 82B44, 82B20, 60K35

1986 *Physics and Astronomy Classification Scheme*: 05.20, 75.40

*Keywords*: Spin-glasses, Sherrington-Kirkpatrick model, Edwards-Anderson model, disordered systems.

Talk presented at the satellite conference to the XIth International congress of mathematical physics, devoted to the 'mathematical physics of disordered systems', held in Paris, 25–27 July 1994.

# 1    Introduction and physical motivations

Equilibrium statistical mechanics of translationally invariant (or periodic) systems is well understood; although particular problems can be very hard or impossible to solve analytically, this discipline provides a scheme for the treatment of problems arising in condensed matter physics on which we can confidently rely both mathematically and numerically. From the mathematical point of view, equilibrium statistical mechanics is a logically closed theory that explains the regularity of thermodynamic quantities and the phenomenon of phase transition; it achieved its ultimate stage thanks to the works of Dobrushin [36] and Lanford and Ruelle [87].

Several physical systems fail however to fulfill the translation invariance (or periodicity) condition; these systems fall into two classes: quasiperiodic (like quasicrystals) and random systems (like spin glasses). Although systems in these two classes share some common features, their treatment is not yet unified. In this review, attention is paid only to random systems; readers interested in quasiperiodic systems may consult [57, 84, 103, 45] for some partial results.

Spin glasses are systems whose translational invariance is broken by the presence of frozen randomness. It is not clear what is meant by 'frozen randomness'; it was originally believed that this randomness can evolve under the dynamics to some non random interaction. Nowadays, it is generally accepted that this randomness is deeply frozen and cannot evolve. From a fundamental point of view, it is however questionable whether equilibrium statistical mechanics is the appropriate framework for their study since these systems are *stricto sensu* not in equilibrium but in some relaxing metastable state. For instance, the ordinary industrial glass is a metastable (but not random) state of silicium dioxide that can be also found in nature under two other stable phases: quartz crystals and sand. Visiting any museum exhibiting objects from the classical antiquity can convince you however that the relaxation time needed for the transformation of glass into one of its stable phases exceeds historical times and so the use of equilibrium statistical mechanics, although it might be only an approximation, is 'ontologically' justified for the study of glasses. The more recent belief is even that the spin glasses are not metastable systems and cannot thermodynamically evolve.

Like glass that are deterministic non translationally invariant deterministic systems, *spin glasses* are non translationally invariant magnetic systems with frozen randomness. After some controversy, in the beginning, about which objects should be designed by the vocable spin glasses, it is generally accepted, nowadays that they fall into three categories [8].

*Non-stoicheiometric alloys:* these are typically alloys composed by a nonmagnetic atom of a noble metal (like gold, platinum, cooper or silver) and a magnetic atom of a transition metal (like iron or manganese) in a proportion not satisfying chemical valence saturation, eg $Au_{1-x}Fe_x$. There is a periodic matrix-crystal of gold but a proportion $x$ of the crystaline sites, randomly scattered through the lattice, are occupied by iron. The magnetic interaction is only between the magnetic (transition) atoms and has the form $J(|\mathbf{r}|) = J_0 \frac{\cos(2k_F|\mathbf{r}|)}{(k_F|\mathbf{r}|)^3}$, where $k_F$ is the Fermi wavenumber, $J_0$ a constant depending on the nature of the metals in the composition of the alloy, and $\mathbf{r}$ the distance between the magnetic atoms. Since the magnetic atoms are not occupying periodically the sites of the nonmagnetic matrix-crystal, the effective interaction has "randomly" alternating signs.

*Random occupation of crystal sites:* these are non stoicheimetric ternary alloys of the form $Eu_xSr_{1-x}S$ (where the sulfur atom can be replaced by selenium or tellurium). Again there is a periodic crystalline structure but the magnetic atoms are randomly scattered through the lattice sites. The magnetic interaction is between europium and/or strontium and is ferromagnetic (positive) when these atoms are neighbours and antiferromagnetic (negative) when they are next nearest neighbours.

*Amorphous structure of material:* these are noncrystalline alloys of the form $Al_{0.63}Gd_{0.37}$ where the atoms of aluminium and gadolinium are at random positions in the space.

All these materials exhibit similar thermodynamic behaviour, *ie* some features are sample independent and some other are sample dependent (random). For instance, in the figure below is plotted the real part of the magnetic susceptibility $\chi$ as a function of temperature $T$ for an $Eu_xSr_{1-x}S$ alloy.

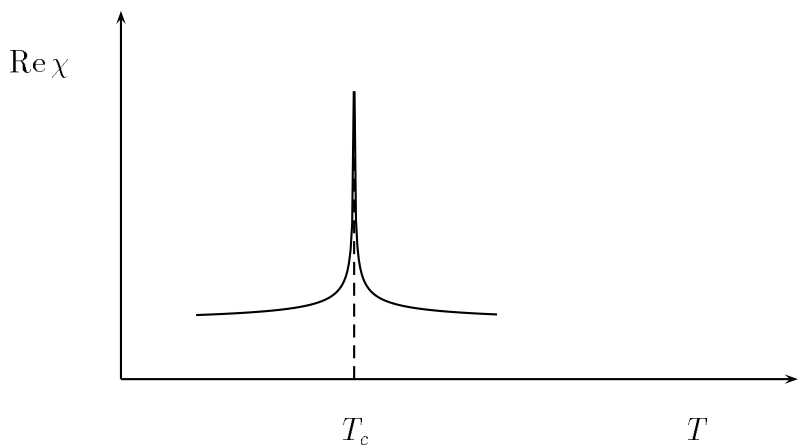

Figure 1: Qualitative behaviour of the real part of the magnetic susceptibility $Re\,\chi$, as a function of temperature $T$. A cusp-like singularity is observed for $T_c$.

This function has a cusp-like singularity at a critical temperature $T_c$ that is sample independent. Similarly, in the following figure is plotted the order parameter $q$ as a function of temperature for an $Al_{0.63}Gd_{0.37}$ alloy.

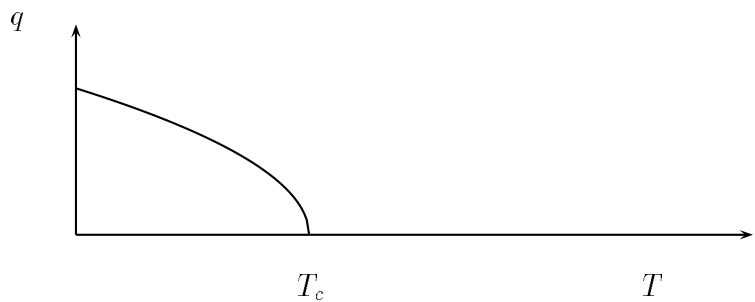

Figure 2: Qualitative behaviour of the order parameter $q$, as a function of temperature $T$. An infinite slope singularity is observed for $T_c$.

Again, the value of the critical temperature is sample independent.

Having in mind the previous observations, it is clear that *any reasonable theoretical model for spin glasses, must include the main structural features of these systems and reproduce the qualitative experimental behaviour.*

## 2 Models proposed for spin glasses and formulation of mathematical problems

In view of the physical properties of spin glasses, it seems reasonable to introduce random interactions. But the real systems are so complex that realistic models of randomness are quite intractable both mathematically and numerically. For this reason — as usual in mathematical physics — some simplified models mimicking the main features of real systems were introduced. Instead of giving a detailed description of all these models, a synoptic table with the relevant references and the main features of most of these models is given below. Neither the table nor the quoted references are exhaustive!

| Model | Type | Relevant references |
|---|---|---|
| Random energy models | U/NA | [33, 125, 55, 107, 119, 120, 82, 85, 21] |
| Spin glasses on Cayley trees | U/NA | [28, 27, 56, 62, 10] |
| Multiplicative chaos and related models | U/NA | [29, 117, 34, 17, 24, 133, 35] |
| Random diluted ferromagnet | U/D | [91, 64, 49, 50, 6, 26, 60, 122, 38] |
| Random field Ising model ($d \geq 2$) | U/D | [73, 72, 15, 2, 50, 47, 9, 23] |
| Pastur-Figotin (Hopfield) model | F/MF | [113, 71, 116, 131, 132, 126, 13, 14, 80, 78, 5] |
| $N$–vector models | F/MF | [80, 43, 44, 121] |
| SG with semiconvergent interactions | F/D | [124, 51, 52, 136, 137, 40, 41, 42, 69, 20, 18, 19, 80] |
| SG on Voronoï lattice or percolation clusters | F/D | [46, 65, 7, 134, 6] |
| Sherrington-Kirkpatrick model | F/MF | [127, 77, 1, 30, 83, 128, 32, 66, 67, 68, 31, 99, 37] |
| Edwards-Anderson model | F/D | [39, 115, 13, 129, 40, 81, 11, 12] |

Models are classified according to two criteria: their range of interaction and their frustration character. Concerning the *range of interactions*, we distinguish three classes (D, MF, and NA); D stands for decaying interactions, *ie* the absolute value of the interaction becomes smaller when the distance between the interacting magnetic atoms becomes larger; MF stands for mean-field interactions where the strength of the interactions keeps the same magnitude all over the sample; finally, NA stands for nonapplicable and it is used for some particular models defined on lattices without natural underlying metric structure (lattices not embeddable into $\mathbb{R}^d$ in a Lipschitz way).

*Frustration* is a phenomenon occurring each time a binary relation that is reflexive and symmetric fails to be transitive, eg friendship is such a relation since (A and B friends) and (B and C friends) does not imply (A and C friends)! In the context of spin glasses, frustration occurs when interactions can take both signs and the spins live on a lattice with loops, like $\mathbb{Z}^d$ for $d \geq 2$. In the previous table, we distinguished two classes U and F according to the fact that the model is unfrustrated or frustrated.

Two of these models, the Sherrington-Kirkpatrick and the Edwards-Anderson models are more precisely defined below and studied in the subsequent sections. Both models are defined on a configuration space $\Sigma_N = \{-1, 1\}^{\Lambda_N}$ over a finite set of sites $\Lambda_N$. Configurations are denoted $\sigma \in \Sigma_N$ and $\sigma_i \in \{-1, 1\}$ denotes the spin value over the site $i \in \Lambda_N$. Eventually, the finite parameter $N$ will be allowed to tend to infinity (thermodynamic limit).

*The Sherrington-Kirkpatrick model:* is a mean-field model defined over the set of sites $\Lambda_N = \{1, \cdots, N\}$. We identify the dual lattice $\Lambda_N^\star$ of $\Lambda_N$ with the complete graph $K_N = \{\{i, j\} : i \in \Lambda_N, j \in \Lambda_N, i \neq j\}$ over $N$ and we consider a family of centered, variance 1, and independent, Gaussian random variables $(J_{ij})_{\{i,j\} \in \Lambda_N^\star}$ indexed by $\Lambda_N^\star$. The Hamiltonian of the model is given by

$$H_N(\sigma) = -\frac{1}{2\sqrt{N}} \sum_{\{i,j\} \in \Lambda_N^\star} J_{ij} \sigma_i \sigma_j.$$

Notice that the sum extends over $|\Lambda_N^\star| = N(N-1)$ terms and that the normalisation is in $\sqrt{N}$ so that the central limit theorem is far from being applicable[2]. The choice of Gaussian

---

[2]Depending on the computation in view, some other form of the Hamiltonian may be more appropriate, like, for instance, the form $H'_N(\sigma) = -\frac{1}{\sqrt{N}} \sum_{1 \leq i < j \leq N} J_{ij} \sigma_i \sigma_j$. The two forms $H$ and $H'$ are thermodynamically

variables is done for computational convenience; it is believed that any symmetric distribution with finite moments should lead to the same behaviour. Since the interactions are random, the Hamiltonian is a random variable of the configurations.

*The Edwards-Anderson model:* is defined on the $d$-dimensional lattice. Consider the finite lattice volume $\Lambda_N = [-N, N]^d \cap \mathbb{Z}^d$. The dual lattice $\Lambda_N^*$ is defined as usual and can be identified with the set $\Lambda_N^* = \{\{i, j\}, i \in \Lambda_N, j \in \mathbb{Z}^d, |i - j| = 1\}$. For a family of centered, variance 1, independent, Gaussian random variables $(J_{ij})_{\{i,j\} \in \Lambda_N^*}$ indexed by $\Lambda_N^*$, the hamiltonian is defined by

$$H_N(\sigma) = - \sum_{\{i,j\} \in \Lambda_N^*} J_{ij} \sigma_i \sigma_j.$$

Obviously, this nearest neighbour Hamiltonian is a random function of the configurations.

For both models, the partition function is defined as usual by

$$Z_N = \sum_{\sigma \in \Sigma_N} \exp(-\beta H_N(\sigma))$$

where the parameter $\beta$ is the inverse temperature and for every fixed $\beta$ it is a random variable.

Similarly, the *quenched free energy* is defined by

$$F_N(\beta) = -\frac{1}{\beta} \log Z_N(\beta)$$

and the quenched specific free energy by

$$f_N(\beta) = \frac{1}{|\Lambda_N|} F_N(\beta).$$

Both quantities are random variables. But contrary to the translation invariant case, taking expectation (average over the random variables $J$) of the partition function before computing the free energy, we can define a new quantity, the *annealed free energy*

$$\overline{F}_N(\beta) = -\frac{1}{\beta} \log \mathbb{E} Z_N(\beta)$$

and the annealed specific free energy

$$\overline{f}_N(\beta) = \frac{1}{|\Lambda_N|} \overline{F}_N(\beta),$$

where $\mathbb{E}(\cdot)$ denotes the average over randomness (*ie* average over the random variables $J$). In all the above definitions, special care has been taken in the signs and the normalisations appearing in various formulae; in particular, the sign and the normalisation of the free energy are those imposed by the laws of thermodynamics. Similarly, the minus sign in the exponent of the Boltmann factor, although irrelevant for probabilistic statements concerning symmetric equivalent; they can be made equivalent in dynamical respects as well if the interaction matrix $\mathbf{J} = (J_{ij})_{ij}$ is choosen symmetric.

distributions, is there to remind the reader that physically the most probable configurations are those minimising the Hamiltonian and not those maximising it!

Thermodynamic averages are computed through the random "Gibbs" measure

$$\mu_{N,\beta}(\sigma) = \frac{\exp(-\beta H_N(\sigma))}{Z_N(\beta)}$$

and are usually denoted, in the physical literature, by $\langle \, \cdot \, \rangle_{N,\beta}$, the precise meaning of this symbol[3] being

$$\langle \, \cdot \, \rangle_{N,\beta} = \int (\cdot) \mu_{N,\beta}(d\sigma).$$

Notice however that for the mean field models, this measure does not give rise to an infinite volume Gibbs measure in the sense of Dobrushin-Lanford-Ruelle but only in the weak sense [4, 13].

For translation invariant systems, a phase transition is characterised by the change of an order parameter that can be chosen to be the magnetisation per site. In disordered systems, this is a random quantity so that an average over randomness must be taken. However, the average over the magnetisation vanishes due to the symmetry of the random variables $J$. Various order parameters have been introduced, starting from the Edwards-Anderson one [39]; the smoother seems to be the one defined in [41] by

$$q_N^2 = \frac{1}{|\Lambda_N|} \sum_{i \in \Lambda_N} \mathbb{E}(\mu_{N,\beta}(\sigma_i)^2).$$

# 3 Heuristic results on the Sherrington-Kirkpatrick model

The study of the Sherrington-Kirkpatrick model proved rather complicated. The main difficulty stems from the nonlinearity of the logarithm function appearing in the expression for the quenched free energy. As a matter of fact, it is an elementary observation that

$$\log Z_N = \lim_{R \to 0} \frac{Z_N^R - 1}{R}.$$

So, in [39] and later in [127, 77], a trick was proposed to overcome this difficulty: instead of computing $\mathbb{E} \log Z_N$, it is advised to compute $\mathbb{E} Z_N^R$ where $R$ is the number of replicas of the model, *ie* independent (in $\sigma$) copies of the model all having *the same* random interactions $J$. So for $R$ a positive integer, this reduces to computation of the moments of the partition function. Eventually, the computations for integer positive values are extended to zero; this is the famous *replica trick*. The first steps of this computation are given below. Fix some positive integer $R$

---

[3]In this paper the symbol $\langle \, \cdot \, \rangle$ is used later to denote the previsible increasing process of the Doob's decomposition of a submartingale. Therefore, we stick to the symbol $\mu$ for Gibbs' averages to avoid any confusion.

and compute

$$
\begin{aligned}
\mathbb{E} Z_N^R &= \mathbb{E}\left([\sum_\sigma \exp(\frac{\beta}{\sqrt{N}}\sum_{i<j} J_{ij}\sigma_i\sigma_j)]^R\right) \\
&= \exp(-\frac{\beta^2 R^2}{4})\sum_{\sigma^1,\cdots,\sigma^R}\exp(\frac{\beta^2}{4N}\sum_{\alpha,\gamma}(\sum_i \sigma_i^\alpha\sigma_i^\gamma)^2),
\end{aligned}
$$

where lower (Roman) indices stand for sites and take values in $\Lambda_N$ and upper (Greek) indices stand for replicated copies and take values in $\{1,\cdots,R\}$.

Now use the elementary identity

$$
\exp(\lambda a^2) = \frac{1}{\sqrt{2\pi}}\int_{-\infty}^\infty \exp(-\frac{Q^2}{2}+\sqrt{2\lambda}aQ)dQ
$$

to linearise the exponent appearing in the integral for $\mathbb{E} Z_N^R$ and write finally

$$
\mathbb{E} Z_N^R = \int \exp(-NA(\mathbf{Q}))\prod_{\alpha\neq\gamma}dQ_{\alpha\gamma}
$$

where $A(\mathbf{Q})$ is an effective free energy given by

$$
A(\mathbf{Q}) = -\frac{R\beta^2}{4}+\frac{1}{4}\sum_{\alpha,\gamma}Q_{\alpha\gamma}^2 - \log Z_{\text{rep}}(\mathbf{Q}),
$$

and where

$$
Z_{\text{rep}}(\mathbf{Q}) = \sum_{s_1,\cdots,s_R}\exp(-\frac{\beta}{2}\sum_{\alpha,\gamma}Q_{\alpha\gamma}s_\alpha s_\gamma)
$$

represents an effective partition function over the replica space. Notice that up to this point, the computations presented are absolutely rigorous.

Assume now that the following, *totally unjustified*, statements are true:

- the limit $R\to 0$ does have a meaning,

- the limits $R\to 0$ and $N\to\infty$ commute,

- all integrals converge in the limit, and

- since the functional $A(\mathbf{Q})$ is invariant under the symmetric group $S_R$ over the replicas, so is the solution $q=\arg\min A(\mathbf{Q})$.

Under these assumptions, it is shown in [127, 77] that in the infinite volume limit, the solution for the free energy is given by

$$
f_\infty = \overline{f}_\infty(1-q)^2 - \frac{1}{\sqrt{2\pi}\beta}\int_{-\infty}^\infty \exp(-z^2/2)\log[2\cosh(\beta\sqrt{q}z)]dz,
$$

where the quantity $q$ plays the *rôle* of an order parameter and is given by the mean field self-consistent equation

$$q = \frac{1}{\sqrt{2\pi}\,\beta} \int_{-\infty}^{\infty} \exp(-z^2/2) \tanh(\beta\sqrt{q}\,z) dz,$$

and $\overline{f}_\infty = -\beta/4$ is the annealed free energy. The implicit equation for $q$ cannot be solved in general; notice however that for small values of $\beta$, actually $\beta < 1$, the only possible solution is $q = 0$. This implies that at high temperature and in the replica approximation scheme the quenched free energy coincides with the annealed free energy. For $\beta > 1$, there is a strictly positive solution implying that, at low temperature, there is a non vanishing value for $q$ so that it can be interpreted as an order parameter; moreover, the annealed and quenched free energies do not coincide. What is remarkable is that these very naive computations reproduce roughly the qualitative behaviour of the model as it can be obtained from computer simulations. Of course, one does not expect that such an approximation may faithfully reproduce the exact quantitative behaviour and as a matter of fact, there is a severe problem with the solution at zero temperature: the value of the free energy predicted by the replica trick at zero temperature violates the laws of thermodynamics since it corresponds to negative entropy for the Sherrington-Kirkpatrick model.

What is even more remarkable is that adding some even more unjustified assumptions than the previous ones, Parisi [108, 109, 110, 111] obtained an even more plausible heuristic solution. The Parisi's *Ansatz* stems from the observation that, as usual in the context of phase transition, there is a breaking of the symmetry group. Since the only symmetry group available here is the full symmetric group $S_R$ over replicas, it is worth breaking it. Assume that the solution $q$ found previously is not a minimum but a saddle point. Making his *Ansatz*, Paris fixes some integer $m$ with $0 < m < R$ that is a divisor of $R$ (mind that eventually $R$ goes to zero!) and searches for minimising solutions of the form

$$Q_{\alpha\gamma} = \left\{ \begin{array}{ll} q_0 & \text{if} \quad [\frac{\alpha}{m}] = [\frac{\gamma}{m}] \\ q_1 & \text{if} \quad [\frac{\alpha}{m}] \neq [\frac{\gamma}{m}], \end{array} \right.$$

where $[\cdot]$ denotes the integer part. Repeat now the computations for the infinite volume free energy. The behaviour predicted now is much more reasonable and there is almost no violation of the thermodynamic laws. Parisi proposed even to continue the replica symmetry breaking to more than one levels. He claims even that eventually, a continuous function $q(x)$ must be introduced to index the minimising solution. This infinite replica symmetry breaking induces an *ultrametric* structure to the space of states [100].

These results not only lack any rigorous justification but even their formulation in mathematical language is problematic. Quite surprisingly, there is an alternative heuristic formulation, known as *cavity method* [101], that predicts similar (non rigorous) results, confirming thus, by another method, the results of Parisi.

In the following figure the low temperature behaviour of the specific free energy for the Sherrington-Kirkpatrick model within one breaking of the replica symmetry is plotted as a function of the temperature.

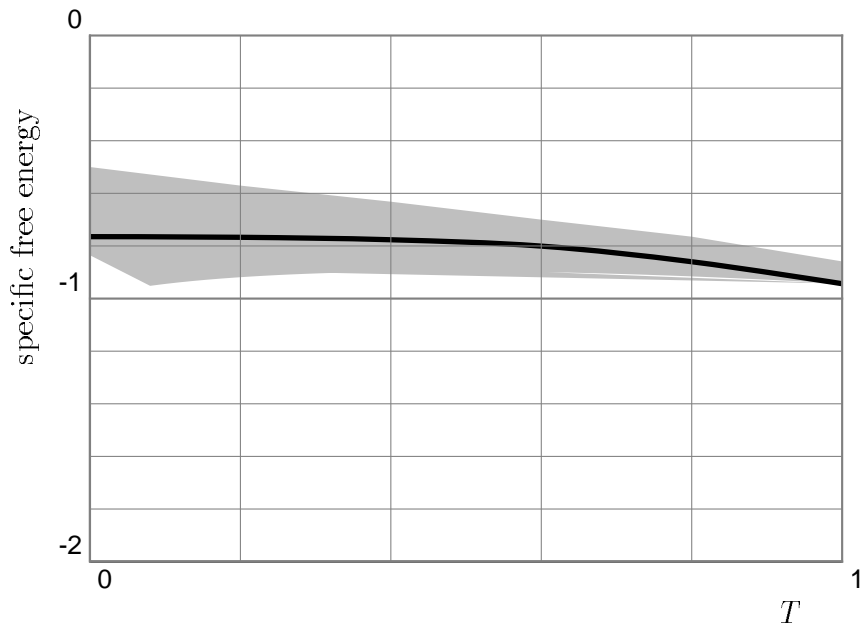

Figure 3: The solid curve represents the value of the specific free energy of the Sherrington-Kirkpatrick model, as predicted by the one replica symmetry breaking computation of [109], as a function of the temperature $T$, for the low temperature regime $T < 1$. The gray region delimits the region where the specific free energy can lie, provided it exists, and it is computed from the rigorous upper bound for $\limsup_N f_N$ of [1] and one of the infinite family of rigorous lower bounds for $\liminf_N f_N$ of [83] best at very low temperature and the rigorous bound of [31] best at intermediate temperatures. No optimisation over the known bounds is performed neither any use of convexity properties is made to delimit the gray area.

The reader can observe how plausible this solution looks compared with the rigorous results.

# 4  Mathematical results for the Sherrington-Kirkpatrick model

In spite of continuing efforts, only partial results are known on this model. The first results were obtained with standard methods of mathematical physics, namely expansions. Almost simultaneously, in 1987, using cluster expansion [51, 53, 54] or graph expansion [1] the high temperature regime was almost completely understood. In [51, 53, 54], various important results concerning the very high temperature region are obtained. It seems to the author that the article [1] is however more complete in the sense that it covers the whole high temperature region and obtains some very partial results in the low temperature regime. In 1993, a totally new approach [30], using stochastic calculus, is introduced to tackle the model. Although no fundamentally new results were obtained with this last method, it has the merit of introducing a totally fresh way of treating the problem. However, both expansion and stochastic methods proved unable, up to the moment these lines are written, to overcome the singularity of $\beta = 1$ that has different origins in the two methods: for expansion methods, it corresponds to the radius of convergence of power series, for stochastic calculus, it stems from the explosion of the autocovariance of the process. Therefore, for the moment, only the high temperature region is accessible. It is not intended to give here a complete report on these two important approaches but only the flavour of the methods and direct the interested reader to the original papers.

The main result of [1] is formulated in terms of the parameter $\tau$ defined by

$$\tau_N(\beta) = \frac{2}{N(N-1)} \sum_{i<j} \mu_{N,\beta}(\sigma_i \sigma_j)^2$$

that is a kind of order parameter for spin correlations.

**Theorem 4.1** *For every $\beta < 1$,*

1. $\lim_{N \to \infty} \mathbb{E}\tau_N(\beta) = 0$,

2. $\lim_{N \to \infty} \frac{1}{N} \mathbb{E} \log Z_N(\beta) = \lim_{N \to \infty} \frac{1}{N} \log \mathbb{E} Z_N(\beta)$,

3. $Z_N/\mathbb{E}Z_N$ *converges in distribution, when $N \to \infty$, to the random variable $\exp(V - v^2/2)$ where $V$ is distributed according to $\mathcal{N}(0, v^2)$ and $v^2 = -[\log(1 - \beta^2) + \beta^2 - \beta^4/4]/2$.*

*Sketch of the proof:* (For the details see [1]). It is enough to prove the convergence in distribution of $Z_N/\mathbb{E}Z_N$ for then the second claim follows immediately and then it is not hard to show that the first claim also holds. Rewrite

$$Z_N = \sum_\sigma \exp[\frac{\beta}{\sqrt{N}} \sum_{i<j} J_{ij} \sigma_i \sigma_j] = (\prod_{i<j} \cosh \frac{\beta J_{ij}}{\sqrt{N}}) \hat{Z}_N,$$

where

$$\hat{Z}_N = \sum_\sigma \prod_{i<j} (1 + \sigma_i \sigma_j \tanh \frac{\beta J_{ij}}{\sqrt{N}}).$$

Expand now the product, assigning a random weight $w_b = \tanh \frac{\beta J_{ij}}{\sqrt{N}}$ to every pair $b = ij$, and perform the sum over $\sigma$. Remark that this sum is symmetric so that only terms where $\sigma_i$'s appear an even number of times, for every $i$, remain. It is convenient to visualise the sum in terms of graphs over the sites. Every vertex is labelled by a site $i$, a bond connects two distinct vertices and carries the corresponding random weight; each vertex $i$ has as many $\sigma_i$'s attached as its graph degree (*ie* the number of bonds emanating from the vertex). Due to the symmetry of the sum, mentioned above, only (simple or multiple) loops remain in this expansion and the quantity $\hat{Z}_N$ can be expanded pictorially as

$$\hat{Z}_N = \quad \cdots \quad + \quad \cdots \quad + \quad \cdots \quad + \ldots \qquad \rbrace \text{ simple loops}$$

$$\cdots \quad + \quad \cdots \quad + \ldots$$

$$\cdots \quad + \quad \cdots \quad + \ldots \qquad \rbrace \text{ multiple loops}$$

Looking at the graphs of this expansion, one observes that the multiple loops appearing there are just the gluing of simple graphs; this is called an exponential family[4] in combinatorial theory [90, 135]. Denoting by $V_N = \sum_{\gamma: \text{simple loops}} w(\gamma)$ the sum of the contributions over the simple loops, then the complete sum can be written $\hat{Z}_N = \exp(V_N - \text{small corrections})$.

The rest of the proof reposes on one simple idea: we are interested on the infinite volume limit; thus provided that taking the limit $N \to \infty$ does not lead out of the convergence domain of the power series, we can start by taking this limit to get the asymptotic behaviour. This idea is used several times in the proof. We illustrate it by proving a very simple intermediate result, namely that $\mathbb{E} V_N^2 = v^2$. In fact, write

$$V_N = \sum_{\gamma: \text{simple loops}} w(\gamma) = \sum_{k \geq 3} \sum_{\{\gamma: |\gamma| = k\}} w(\gamma)$$

where we have splitted the sum over simple loops into a sum of all possible lengths of simple loops (hence the condition $k \geq 3$) and into a sum over simple loops of given length. Now $\tanh(\cdot)$ is an odd function and the variables $J_{ij}$ are symmetric; hence the random variables $w(\gamma)w(\gamma')$ are orthogonal if the loops $\gamma$ and $\gamma'$ differ by at least one bond. Therefore,

$$
\begin{aligned}
\mathbb{E} V_N^2 &= \sum_{k \geq 3} \sum_{\{\gamma: |\gamma| = k\}} \mathbb{E} w(\gamma)^2 \\
&= \sum_{k \geq 3} [\mathbb{E}(\tanh \frac{\beta J_{12}}{\sqrt{N}})^2]^k \frac{N(N-1)\cdots(N-k+1)}{2k} \\
&\to \sum_{k \geq 3} \frac{\beta^{2k}}{2k} \\
&= -\frac{1}{2} \log(1 - \beta^2) - \frac{\beta^2}{2} - \frac{\beta^4}{4}.
\end{aligned}
$$

---

[4]As a matter of fact this is not exactly an exponential family since no double bond is allowed in the multiple loops; this constraint introduces a slight correction in the exponentiation formula for the generating function.

The passage to the second line in the above formula is justified by the independence and identical distribution of the random variables $(J_{ij})$ and by a simple combinatorial argument counting the number of length $k$ simple loops over $N$ possible sites: fixing a vertex of the loop that should be called "the first vertex" and a sense of rotation, there are $N$ possible sites that can give their label to the first vertex, $N-1$ for the second vertex until the exhaustion of the loop (hence the numerator). Now there are 2 possible rotation directions and $k$ possible first vertices (hence the denominator).

Arguments of the same kind are then used to show that $V_N$ tends, in distribution to a centered Gaussian random variable of variance $v^2$ and finally prove the theorem. $\square$

This expansion method is powerful and rigorous inside the convergence domain of the power series. It looks quite natural for a physicist and uses only elementary mathematics. On the counterpart, it can be very cumbersome.

The same authors obtain also some low temperature results and particularly the following

**Theorem 4.2** *For $\beta$ sufficiently larger than 1,*

$$\liminf_{N \to \infty} \tau_N(\beta) = 1 - \mathcal{O}(1/\beta).$$

This proves that if the limit exist, $\tau$ is an order parameter for the model and the different values taken in low and high temperatures indicate the existence of a phase transition.

We come now to the result proved by the method of [30]. The main idea is the observation that $Z_N/\mathbb{E}Z_N$ can be expressed as an exponential martingale in the parameter $\beta^2$. The idea to use martingales in the context of statistical mechanics is, quite surprisingly, in a paper that does not deal with statistical mechanics at all but with a simple model for turbulence [74]; this model was extended in [29] to include temperature and in this latter paper, the ratio $Z_N/\mathbb{E}Z_N$ was expressed as a martingale in $N$. Now when a martingale in the volume appears, the martingale convergence theorem gives immediately the thermodynamic limit. Here the martingale character only serves to guess the correct form of the limiting behaviour; the study of the thermodynamic limits necessitating a more detailed treatment.

The starting point is the Hamiltonian of the Sherrington-Kirkpatrick model, where the inverse temperature is incorporated into the Hamiltonian,

$$H_N(\beta; \sigma) = \beta \sum_{i<j} \frac{J_{ij}}{\sqrt{N}} \sigma_i \sigma_j.$$

Since the random variables are distributed according to $\mathcal{N}(0,1)$, the above Hamiltonian is a Gaussian process indexed by the configurations whose covariance is given by

$$\mathbb{E}[H_N(\beta; \sigma) H_N(\beta; \sigma')] = \frac{\beta^2}{N} \sum_{i<j} (\sigma_i \sigma_i')(\sigma_j \sigma_j').$$

Introduce now a family of independent standard Brownian motions $(B_{ij})$ indexed by the bonds $ij$ and define a modified Hamiltonian by

$$\tilde{H}_N(t;\sigma) = \sum_{i<j} \frac{B_{ij}(t)}{\sqrt{N}}\sigma_i\sigma_j.$$

Computing the covariance matrix for the modified process, we find

$$\mathbb{E}[\tilde{H}_N(t;\sigma)\tilde{H}_N(t;\sigma')] = \frac{t}{N}\sum_{i<j}(\sigma_i\sigma_i')(\sigma_j\sigma_j').$$

Therefore, choosing $t = \beta^2$ the two processes are indistinguishable. Denoting by

$$\mathcal{F}_t = \sigma\{B_{ij}(s); 1 \le i < j \le N; s \le t\},$$

we remark that $\tilde{H}_N(t)$ is a square integrable martingale with respect to $\mathcal{F}_t$ and therefore, for every fixed configuration $\sigma$, $\exp[\tilde{H}_N(t) - \langle\,\tilde{H}_N\,\rangle(t)/2]$ is also a martingale with respect to the same $\sigma$-algebra, $\langle\,\tilde{H}_N\,\rangle(t)$ being the compensating process of the submartingale $\tilde{H}_N^2(t)$. The important point is that the martingale character remains valid for the sum over configurations, so that

$$K_N(t) = \frac{1}{2^N}\sum_\sigma \exp[\tilde{H}_N(t) - \frac{\langle\,\tilde{H}_N\,\rangle(t)}{2}]$$

is a martingale having the same distribution with the random variable $Z_N/\mathbb{E}Z_N$.

**Theorem 4.3** *For $t \in [0,1[$, the random process $K_N(t)$ converges in distribution to the process* $\exp(M(t) - \frac{\phi(t)}{2})$, *where* $\phi(t) = \frac{1}{2}\log(\frac{1}{1-t} - t)$, *and $M(\cdot)$ is a centered Gaussian process with independent increments and such that, for $0 \le s \le t \le 1$,*

$$\mathbb{E}[(M_\infty(t) - M_\infty(s))^2] = \exp(\phi(t) - \phi(s)).$$

*Sketch of the proof:* (For details see [30]). Remark that

$$\lim_{N\to\infty} \uparrow \mathbb{E}(K_N(t)^2) = \begin{cases} \exp(\phi(t)) & \text{si } t < 1 \\ \infty & \text{si } t \ge 1. \end{cases}$$

The martingale $K_N(t)$ is in fact an exponential martingale that can be written in terms of another stochastic process $M_N(t)$ as

$$K_N(t) = \exp\left(M_N(t) - \frac{\langle\,M_N\,\rangle(t)}{2}\right).$$

It turns out that $K_N(t)$ is a local martingale that solves the stochastic differential equation

$$dK_N(t) = K_N(t)dM_N(t)$$

and whose quadratic variation verifies the differential inequality

$$\frac{d}{dt}\langle\,M_N\,\rangle(t) \le \frac{N-1}{2}.$$

Integrating this inequality, one obviously obtains that $M_N(t)$ is an $\mathbb{L}^2$-martingale with $\mathbb{E}M_N^2(t) \leq t(N-1)/2$. The most technical part of the proof consists in showing that $\langle M_N \rangle(t)$ converges in probability, as $N \to \infty$, and for $t < 1$ to a deterministic function $\phi$. First remark that $\langle M_N \rangle(t)$ is a strictly increasing process going to $+\infty$ as $t \to \infty$. Therefore, there exists a standard Brownian motion $b$ on $\mathbb{R}^+$ such that $M_N(t)$ can be represented as the value of $b$ at the random time $\langle M_N \rangle(t)$, $ie$

$$M_N(t) = b(\langle M_N \rangle(t)).$$

Next remark that for every $a > 0$ and every $\epsilon > 0$,

$$\mathbb{P}(\{-b(t) \leq a + \epsilon t/2, \forall t \geq 0\}) \geq 1 - \exp(-a\epsilon).$$

The proof of this technical step is achieved by establishing that for every $T < 1$, every $a > 0$, and every $\epsilon > 0$,

$$\lim_{N \to \infty} \mathbb{E}(\mathbb{1}_{A_{a,\epsilon}^N} \sup_{0 \leq t \leq T} |F_\epsilon(\langle M_N \rangle(t) - \phi(t)|) = 0,$$

where $A_{a,\epsilon}^N = \{-M_N(t) \leq a + \epsilon\langle M_N \rangle(t)/2\}$ and $F_\epsilon(x) = [1 - \exp(-(1 + \epsilon)x)]/(1 + \epsilon)$. Since $\mathbb{P}(A_{a,\epsilon}^N)$ can be chosen arbitrarily close to 1, this proves the convergence in probability of $\langle M_N \rangle(t)$ to the deterministic function $\phi$.

Using now this fact and the Rebolledo theorem, it is shown that for $t \in [0, 1[$, the martingale $M_N(\cdot)$ converges in distribution to a centered Gaussian process $M_\infty(\cdot)$ with independent increments such that for $0 \leq s \leq t \leq 1$,

$$\mathbb{E}[(M_\infty(t) - M_\infty(s))^2] = \exp(\phi(t) - \phi(s)).$$

Therefore, the random process $K_N(t)$ converges in distribution, for $t \in [0, 1[$, to the process

$$\exp(M_\infty(t) - \frac{\phi(t)}{2})$$

and thus the main theorem of [1] is recovered by using purely probabilistic arguments. □

The other mathematical result known about the Sherrington-Kirkpatrick model concerns the weak self-averaging property of the specific free energy. *Self-averaging* is a very important property of thermodynamic functions; when it is true, it states that the corresponding function is a trivial random variable in the sense that it does not fluctuate from sample to sample. The first result, valid on the whole region of temperature, was established by Pastur and Shcherbina in a weak sense in [115]: they proved the following concentration result:

**Theorem 4.4** *For every $\beta \geq 0$,*

$$\lim_{N \to \infty} \mathbb{E}[f_N - \mathbb{E}f_N]^2 = 0.$$

Notice that this theorem does not guarantee the existence of the thermodynamic limit, $\lim_N \mathbb{E}f_N$, which may not exist in low temperature. But provided that the limit of this expectation exists, the previous result says that the quadratic fluctuations vanish at the thermodynamic limit. Their method was subsequently applied to the Hopfield model and finer and finer results were obtained as time passed [131, 126, 13]. Moreover, the more recent result of [13] applies to both the Hopfield and the Sherrington-Kirkpatrick model to yield the following

**Theorem 4.5** *For every $\beta \geq 0$ and every $z > 0$,*

$$\mathbb{P}(|f_N - \mathbb{E}f_N| \geq z) \leq 2 \exp(-\frac{z^2 N}{5}).$$

Notice that even this strongest version of weak self-averaging does not guarantee the existence of $\lim_{N \to \infty} \mathbb{E}f_N$, but provided this limit exists, use of the Borel-Cantelli lemma suffices to show that the specific free energy is almost surely a self-averaging quantity.

*Sketch of the proof of theorem 4.5:* First, order the bonds in $\Lambda_N^*$ according to some arbitrary order (eg lexicographic) and write $H_N(\sigma) = -\frac{1}{\sqrt{N}} \sum_{B \in \Lambda_N^*} J_B \sigma_B$ where for a bond $B = ij$ in $\Lambda_N^*$ we denote by $J_B \equiv J_{ij}$ and $\sigma_B \equiv \sigma_i \sigma_j$. Now fix some bond $K \in \{1, \cdots, |\Lambda_N^*|\}$ and a parameter $t \in [0, 1]$ and write a modified Hamiltonian

$$H_N(\sigma; t, K) = -\frac{1}{\sqrt{N}} \sum_{B \neq K} J_B \sigma_B - \frac{t}{\sqrt{N}} J_K \sigma_K.$$

Since the Hamiltonian is now depending on the two additional parameters $t$ and $K$, all the thermodynamic functions, and the free energy $F_N(t, K)$ in particular, depend also on these two parameters. Notice that $F_N(1, K) \equiv F_N$. Compute now the derivative with respect to $t$:

$$\frac{dF_N(t, K)}{dt} = \frac{J_K}{\sqrt{N}} \sum_{\sigma} \sigma_K \frac{\exp[-\beta H_N(\sigma; t, K)]}{Z_N(t, K)}.$$

Observe that the sum over $\sigma$ is just the Gibbs average of the values of spin over the ends of the bond $K$ and hence takes values in $[-1, 1]$; therefore

$$|\frac{dF_N(t, K)}{dt}| \leq \frac{|J_K|}{\sqrt{N}}.$$

Use an equality from elementary calculus to write

$$F_N \equiv F_N(1, K) = F_N(0, K) + \int_0^1 \frac{dF_N(t, K)}{dt} dt$$

and denote by $\mathcal{F}_K = \sigma\{J_1, \cdots, J_K\}$. Now,

$$\begin{aligned}
\psi_K &= \mathbb{E}(F_N | \mathcal{F}_K) - \mathbb{E}(F_N | \mathcal{F}_{K-1}) \\
&= \mathbb{E}(F_N(0, K) | \mathcal{F}_K) - \mathbb{E}(F_N(0, K) | \mathcal{F}_{K-1}) \\
&\quad + \int_0^1 [\mathbb{E}(F_N'(t, K) | \mathcal{F}_K) - \mathbb{E}(F_N'(t, K) | \mathcal{F}_{K-1})] dt.
\end{aligned}$$

The crucial points are the use of the previous bound on the derivative of the free energy and the remark that, in the expression for $F_N(0, K)$, what is measurable with respect to the $\sigma$-algebra $\mathcal{F}_K$ is also measurable with respect to $\mathcal{F}_{K-1}$. Finally, since $F_N - \mathbb{E}F_N = \sum_K \psi_K$, and $|\psi_K| \leq \frac{1}{\sqrt{N}}(|J_K| + \mathbb{E}|J_K|)$ the result is obtained by optimising the Markov inequality. □

In parallel to these major results concerning the model there exist also some partial results concerning various quantities. Let us mention the result on the *supremum* of the Hamiltonian over configurations obtained in [128] that states:

$$0 < C_1 \le \liminf_N \frac{\sup_\sigma H_N(\sigma)}{N} \le \limsup_N \frac{\sup_\sigma H_N(\sigma)}{N} \le C_2 < \sqrt{\log 2}.$$

This result is weaker than the corresponding result contained in equations 2.20 and 2.23 of [1]. Its interest stems in the method with which it was obtained: a comparison of Gaussian processes by Slepian's lemma [88] and a Gilbert-Varshamov bound [92] from information theory are used. In my opinion this is an interesting direction to search for obtaining new results in the spin glass models. In particular, the behaviour of Gaussian processes indexed by the spin configurations, in terms of geometrical properties of the configuration space, must be understood.

Other results concern bounds for the *limes infimum* and *limes superior* of the specific free energy valid in the low temperature region ($\beta > 1$). These results are scattered through four different sources [1, 83, 31, 99] that are presented here as a single theorem:

**Theorem 4.6** *For all $\beta > 1$, we have, in distribution, the following inequalities*

$$-g(\beta) \le \liminf_N f_N \le \limsup_N f_N \le -h(\beta)$$

*where $g(\beta) = g_1(\beta) \wedge g_2(\beta)$ and*

$$g_1(\beta) = \sqrt{\log 2} + \frac{\log 2}{\beta},$$

$$g_2(\beta) = 1 + \frac{\log 2}{\beta} - \frac{\log \beta}{2\beta} - \frac{3}{4\beta}$$

*and*

$$h(\beta) = -\frac{\log 2}{\beta} - \frac{1}{\beta} \int_0^1 \log \cosh(\beta \sqrt{\frac{2u}{\pi}}) du.$$

In spite of the efforts to prove the existence of the specific free energy for this model up to now, such a proof is still missing nowadays. But if this limit exists, it must necessarily be a trivial random variable in the sense that it coincides with its expectation as it is shown in [13]. The bounds obtained in this theorem are used to delimit the gray region of figure 3; as a matter of fact, there is an infinite family of available bounds for the very low temperature regime and $g_1$ must be optimised over all these bounds. Not to burden the presentation, I give here only the easier linear bound for $g_1$.

*Proof of the theorem:* The simplest bound $g_1$ from those in [83] is obtained as follows: denote by $A_N(\sigma)$ the configuration dependent event

$$A_N(\sigma) = \{\exp(\frac{\beta_1}{\sqrt{N}} \sum_{i<j} J_{ij} \sigma_i \sigma_j) \le \exp(\frac{\beta_1^2}{2}(N-1))\}.$$

Then by Chebyshev inequality, the probability of the configuration independent event $B_N = \cup_\sigma A_N(\sigma)^c$ tends to zero, provided that $\beta_1 > 2\sqrt{\log 2}$, hence $B_N^c$ has full measure. On the set $B_N^c$ we have,

$$Z_N(\beta) = \sum_\sigma \exp(\frac{\beta}{\sqrt{N}} \sum_{i<j} J_{ij}\sigma_i\sigma_j) \leq 2^N \exp(\frac{\beta\beta_1}{2}(N-1))$$

and therefore

$$g_1(\beta) = \inf_{\beta_1 > 2\sqrt{\log 2}} (\frac{\beta_1}{2} + \frac{\log 2}{\beta}) = \sqrt{\log 2} + \frac{\log 2}{\beta}.$$

The bound for $g_2$ is obtained in [31], some ideas are also in [99, 61]. The specific free energy of the model is not changed by adding a finite, configuration independent term in the Hamiltonian. Thus, we use the Hamiltonian $H_N$, defined on $\Sigma_N = \{-1, 1\}^N$, given by the formula

$$H_N(\sigma) = -\frac{1}{\sqrt{N}} \sum_{i<j} J_{ij}\sigma_i\sigma_j - \frac{1}{2\sqrt{N}} \sum_i J_{ii}\sigma_i^2,$$

where $(J_{ii})$ is a family of independent $\mathcal{N}(0, 1)$ random variables. This Hamiltonian leads to the same specific free energy (mind that $\sigma_i^2 = 1$ hence the extraneous term is configuration independent and $\frac{1}{\sqrt{N}} \sum_i J_{ii} = \mathcal{N}(0, 1)$ in distribution.)

Let $\mathbf{M} = (M_{ij})$ be a symmetric $N \times N$ matrix with

$$M_{ij} = \begin{cases} \frac{1}{2\sqrt{N}} J_{ij} & \text{if } i < j \text{ or } i > j \\ \frac{1}{\sqrt{2N}} J_{ii} & \text{if } i = j. \end{cases}$$

This matrix has almost surely simple eigenvalues $\lambda_1 < \cdots \lambda_N$ with normalised eigenvectors $\phi^1, \cdots, \phi^N$. Since the distribution of $\mathbb{M}$ is invariant under orthogonal transformations, the diagonal matrix

$$\Lambda = \begin{pmatrix} \lambda_1 & & \\ & \ddots & \\ & & \lambda_N \end{pmatrix}$$

is independent of the orthogonal matrix $(\phi^1, \cdots, \phi^N)$ and we may choose the frame $\phi$ such that it should be uniformly distributed on the set $O(N)$ of orthogonal matrices. In particular, for every positive measurable function $F$, we have

$$\mathbb{E}[F(\mathbf{M})|\Lambda] = \int F(\phi\Lambda\phi^T)\kappa(d\phi) \quad \mathbb{P}\text{-a.s.}$$

where $\kappa$ is the uniform probability on $O(N)$. Change now the configuration space $\Sigma_N = \{-1, 1\}^N$ into the sphere $S_N = \{s \in \mathbb{R}^N : |s|^2 = \sum_{i=1}^N s_i^2 = N\}$ and define a new model (termed spherical model) with the same functional form for the Hamiltonian but defined over the configuration space $S_N$. Its partition function is given by

$$Z_N^{\text{sph}} = \int \exp(-\beta H_N(s))\nu_N(ds),$$

where $\nu_N$ is the uniform probability on the sphere $S_N$.

The main point is the domination of the Sherrington-Kirkpatrick model by the spherical model; for any fixed configuration $\sigma \in \Sigma_N$, the distribution of the scalar products $(\sigma, \phi^i)$ under $\kappa$ is $\nu$ and hence,

$$\mathbb{E}(Z_N | \Lambda) = Z_N^{\text{sph}}.$$

Now this model is soluble and its specific free energy is explicitly bounded by $-g_2(\beta)$ where $g_2(\beta) = -1 - \log 2/\beta + \log \beta/2\beta + 3/4\beta$.

It remains now to show the upper bound, obtained in [1, 99]. Start from a sequential algorithm for obtaining configurations that optimise this upper bound. For a fixed set of couplings $(J_{ij})$, associate with each spin configuration $\sigma$ a new configuration $\eta$ by

$$\eta_1(\sigma, J) = \sigma_1$$
$$\eta_j(\sigma, J) = \begin{cases} \sigma_j \operatorname{sgn} \sum_{i=1}^{j-1} J_{ij}\sigma_i & \text{if} \quad \sum_{i=1}^{j-1} J_{ij}\sigma_i \neq 0 \\ \sigma_1 \sigma_j & \text{otherwise.} \end{cases}$$

The Hamiltonian can now be expressed as

$$H_N(\sigma) = -\sum A_k(\sigma, J)\eta_k(\sigma, J)$$

with $A_k(\sigma, J) = \left| \sum_{i=1}^{k-1} \frac{J_{ij}}{\sqrt{N}} \sigma_i \right|$. Define $\mathcal{F}_k = \sigma\{J_{ij}; i < j \leq k\}$ and remark that conditionally to $\mathcal{F}_k$, the sum $\sum_{i=1}^{k-1} J_{ij}\sigma_i$ follows a $\mathcal{N}(0, k-1)$ law, hence conditionally to $\mathcal{F}_k$,

$$\frac{\mathbb{P}(A_k \in da)}{da} = \begin{cases} \frac{2}{\sqrt{2\pi \frac{k-1}{N}}} \exp(-\frac{N}{k-1} \frac{a^2}{2}) & \text{for} \quad a \geq 0 \\ 0 & \text{for} \quad a < 0. \end{cases}$$

Moreover, for fixed couplings $(J_{ij})$, remark that the map $\sigma \mapsto \eta$ is invertible and denote by $E(\eta) = -\sum_{k=1}^{N} \alpha_k^N \eta_k$ where $\alpha_k^N = \mathbb{E}A_k \simeq \sqrt{\frac{2}{\pi} \frac{k-1}{N}}$. Use now a standard variational principle to express

$$\log Z_N(\beta) = \sup\{-\sum_\sigma \rho(\sigma)\log\rho(\sigma) - \beta \sum_\sigma \rho(\sigma)H_N(\sigma)\}$$

the *supremum* being over probability measures over the configurations. For a given set $(J_{ij})$, consider $\tilde{\rho}(\sigma) = \frac{\exp(-\beta E(\eta(\sigma, J)))}{Z_N^*}$ where

$$Z_N^* = \sum_\sigma \exp(-\beta E(\eta(\sigma, J))) = \sum_\eta \exp(-\beta E(\eta)) = 2^N \prod_{k=1}^{N} \cosh(\beta \alpha_K^N).$$

Therefore,

$$\frac{1}{N} \log Z_N \geq \frac{1}{N} \log Z_N^* - Q_N$$

where

$$Q_N(\beta) = \frac{\beta}{N} \sum_\eta \frac{\exp(-\beta E_N(\eta))}{Z_N^*(\beta)} [H_N(\sigma(\eta, J)) - E_N(\eta)].$$

It is quite straightforward to show (see lemma 5.1 of [1]) that $\lim_N Q_N = 0$ and on the other hand to estimate the value of $\frac{\log Z_N}{N}$ by elementary methods, hence the explicit form for $h$ given in the statement of the theorem immediately follows. $\qquad \square$

Finally, let us mention the results in [66, 67, 68] where only elementary probabilistic methods like Chebyshev inequalities are used to obtain certain high temperature results and where mathematical considerations in combination with some *Ansätze* are used to recover the Parisi solution.

# 5   Mathematical results for the Edwards-Anderson model

Contrary to the Sherrington-Kirkpatrick model, that is a mean field model, the Edwards-Anderson spin glass is a short range system; as such it is believed to modelise more faithfully the physical reality than the previous model. On the counterpart, some of its features are more complicated. Therefore much fewer rigorous results are known.

The first result is the following

**Theorem   5.1** *For every $\beta \geq 0$ and every dimension $d$, the*

$$\lim_{N \to \infty} f_N = f_\infty$$

*exists and is non random.*

*Proof:* an exercise on the law of large numbers! If you are lazy enough to do it, look at [130, 63].
□

All the other features of the model are much more complicated. Especially the structure of the Gibbs states is still very controversial. Namely, there are two different ways arguing: the tenants of the Parisi's way of thinking [102, 94, 95], based on the analogy with the Sherrington-Kirkpatrick model and on numerical simulations, claim that, at low temperature, the Edwards-Anderson model has infinitely many pure states like a genuine spin glass. On the other hand, the followers of the Fisher's and Huse's way of thinking, based on scaling arguments, claim that there are at most two pure phases at low temperature connected by symmetry [48]. The rigorous results obtained so far [40, 81] do not allow to settle the controversy.

It seems that one of the problems of the model is that, contrary to mean field systems, the annealed free energy is not a good starting point, even at high temperature, in the sense of the following

**Theorem   5.2** *For every $\beta > 0$ and every dimension $d$, the strict inequality*

$$f_\infty > \overline{f}_\infty$$

*holds.*

A non strict inequality immediately follows from the concavity of the logarithm function and the Jensen inequality. Some indication in the direction of strict inequality is given in [59] and this fact was proven in [81] for Gaussian variables and in [32] for $\pm 1$ Bernoulli variables. Here an elegant unpublished proof, due to Varadhan [129], valid for Gaussian variables is given.

*Proof:* Instead of using the usual definition for the partition function, use a normalised partition function

$$Z_N = \frac{1}{2^{|\Lambda_N|}} \sum_\sigma \exp(\beta \sum J_{ij} \sigma_i \sigma_j).$$

This definition introduces an irrelevant constant shift of the free energy but has the advantage to replace the counting measure over configurations by a uniform probability on the configuration space. Consider the set

$$A_\epsilon = \{\frac{1}{2d|\Lambda_N|} \sum_{\{i,j\} \in \Lambda_N^\star} J_{ij}^2 \le 1 + \epsilon\}.$$

Using Markov inequality, it is easy to show that for every $\epsilon > 0$, we have $\lim_N \mathbb{P}(A_\epsilon^c) = 0$ and hence that the set $A_\epsilon$ has full measure in the thermodynamic limit. Using this remark, we can show that

$$\lim_N \frac{1}{|\Lambda_N|} \mathbb{E} \log Z_N = \lim_N \frac{1}{|\Lambda_N|} \mathbb{E}(\mathbb{1}_{A_\epsilon} \log Z_N).$$

Using the trivial identity

$$1 = \exp(-\frac{\lambda}{2} \sum J_{ij}^2) \exp(\frac{\lambda}{2} \sum J_{ij}^2)$$

we can compute

$$
\begin{aligned}
\frac{1}{|\Lambda_N|} \mathbb{E}(\mathbb{1}_{A_\epsilon} \log Z_N) &= \frac{1}{|\Lambda_N|} \mathbb{E}(\mathbb{1}_{A_\epsilon} \log[\frac{1}{2^{|\Lambda_N|}} \sum_\sigma \exp(\beta \sum J_{ij} \sigma_i \sigma_j)]) \\
&\le \frac{1}{|\Lambda_N|} \mathbb{E}\left(\mathbb{1}_{A_\epsilon} \log[\frac{1}{2^{|\Lambda_N|}} \sum_\sigma \exp(\beta \sum J_{ij} \sigma_i \sigma_j - \frac{\lambda}{2} \sum J_{ij}^2) \exp(\frac{\lambda}{2} |\Lambda_N|(1+\epsilon))]\right)
\end{aligned}
$$

Now Jensen's inequality can be used, explicit integration over the Gaussian variables performed, the limit $\epsilon \to 0$ taken, and the optimisation with respect to the parameter $\lambda$ done, to obtain, on the set $A_\epsilon$,

$$
\begin{aligned}
\lim_N \frac{1}{|\Lambda_N|} \mathbb{E} \log Z_N &\le \frac{d}{2}[\sqrt{1+4\beta^2} - 1 - \log(\frac{1+\sqrt{1+4\beta^2}}{2})] \\
&< d\beta^2 = \lim_N \frac{1}{|\Lambda_N|} \log \mathbb{E} Z_N.
\end{aligned}
$$

$\square$

# 6 Other results on the Edwards-Anderson model

Besides the mathematical results presented previously, there are also extensive numerical simulations on this or other, closely related models, that aim to prove (or disprove) the claims of the tenants of the Parisi's school. The early computer simulations were quite imprecise since they largely underestimated the relaxation phenomena [106]. Recently, much more precise large scale simulation are performed [22, 93].

The intuitive picture Parisi has of the low temperature phase structure is that there are several pure phases, denoted by Greek indices in the sequel. Each phase $\alpha$ corresponds to an extremal Gibbs state (*ie* a measure with respect to which thermal averages $\mu_\alpha(\cdot)$ are computed). An interesting quantity to compute is the overlap parameter $q_{\alpha\gamma} = \frac{1}{N}\sum_{i=1}^{N} \mu_\alpha(\sigma_i)\mu_\gamma(\sigma_i)$. In the low temperature phase, there is a mixture of such pure phases, each phase contributing with a weight $W^\alpha$. For every realisation of the randomness, compute the probability distribution

$$P_N^J(dq) = \mathbb{P}(q_{\alpha\gamma} \in [q, q+dq]) \rightarrow \sum_{\alpha,\gamma} W_\alpha W_\gamma \delta(q_{\alpha\gamma} - q)dq.$$

The important point is that the weights $W$ are supposed to be random variables, even in the thermodynamic limit, *ie* they are not self averaged (sample dependent) quantities. This belief is in the crux of the Parisi's solution and is slightly supported by the numerical evidence. Taking an average of the probability density for $q_{\alpha\gamma}$ over various realisations of the randomness, a probability measure having continuous distribution down to 0 is numerically obtained. In the figure below, numerical results on the averaged probability distribution are quoted from [22].

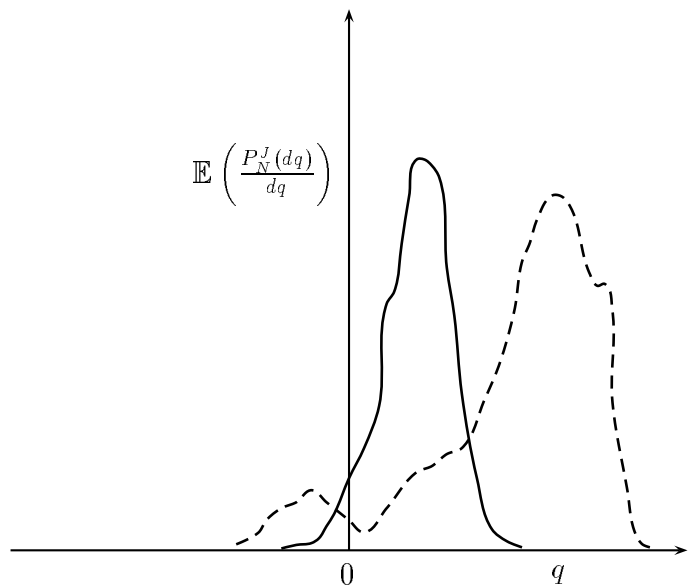

Figure 4: This figure represents the averaged over the randomness probability density for $q$. The solid line represents the high temperature ($T > 1$) distribution and the dashed one the low temperature ($T < 1$) distribution. The results are from [22] and correspond to several hundreds of hours of computer time of Monte Carlo simulations on a finite systems supposed to be so large that the thermodynamic limit is attained. What is meant by 'large' depends on the computer budget your laboratory disposes and in the present case it means exactly a box of $15 \times 15 \times 15$ sites. We remark that the high temperature distribution is — quite astonishingly — not symmetric around 0 and that the low temperature one develops a small but strictly positive density in the negative axis. This may be due to the small size of the system that feels the boundary condition. In spite of the enormous amount of computer time needed to produce such a figure, in contrast to similar simulations on deterministic systems, it only qualitatively suggests that there is a phase transition.

The continuous distribution extending down to 0 is interpreted as the lack of self-averagness.

In a recent paper by Newman and Stein [105] it was argued that this picture cannot be correct for the Edwards-Anderson model. These authors proved that if the $\lim_N P_N^J$ exists *before* averaging over realisations of $J$'s (as it is claimed in the Parisi's conjectures), then it is automatically self-averaged. Now the numerical evidence and the Parisi's claims seem to show that "$\lim_N$" $P_N^J$ is not self-averaged. Therefore, if this quantity is not self-averaged it merely does not exist (*ie* the thermodynamic limit does not exist). If the limit does not exist this is the signature of a chaotic dependence of $P_N^J$ that is consistent only with an infinite range model. So, it is claimed that this picture of infinitely many low temperature pure phases is inconsistent with the finite range of interactions of the Edwards-Anderson model. So as time passes, it seems that the nature of the low temperature phases of the model gets more and more controversial.

# 7  Conclusion

I should like to close this survey at this controversial point. It is not possible in the limited space of this survey to treat very interesting related problems. For instance the more realistic quantum case is completely left out of this review.

It is shown in the introduction of this paper that the disordered systems might be thermodynamically out of equilibrium. An interesting fundamental question should be to ask how these systems evolve in time, how the equilibrium is reached and what are the mechanisms of the the metastability that has been observed experimentally. Some of these questions have started to be adressed nowadays, especially using numerical methods. The mathematically rigorous study of dynamics that became quite sophisticated for deterministic systems [79, 96, 97, 98] is still in en embryonic stage of development.

Important connections of the statistical mechanics of disordered systems with the spectral theory of random operators [114] are also almost absent from this paper. One reason is that there is an excellent review by Pastur [112] dealing with these topics. The other reason is that such an exposition of this direction, should invariably lead us to another closely related *terra incongnita*, namely random walks in random environments that deserves a review by its own (see [123, 16]).

Finally, no account of recent developments in the theory of stochastic simulations is given. For an introduction to the topic one should look at [118] and for more specific developments in the context of spin glasses, one can consults the more specialised articles [86, 22].

Fresh results and theoretical advancements in this fascinating domain of spin glasses are still needed.

**Acknowledgments:** The author profited from numerous useful discussions with several colleagues. He acknowledges particularly the discussions he had with F Koukiou, J Mémin, and N Sourlas.